# Melanin-Based Compounds as Low-Cost Sensors for Nitroaromatics: Theoretical Insights on Molecular Interactions and Optoelectronic Responses

*Published as part of ACS Omega special issue "Chemistry in Brazil: Advancing through Open Science".*

João P. Cachaneski-Lopes, Felipe Hawthorne, Cristiano F. Woellner, Toby L. Nelson, Roger C. Hiorns, Carlos F. O. Graeff, Didier Bégué, and Augusto Batagin-Neto*

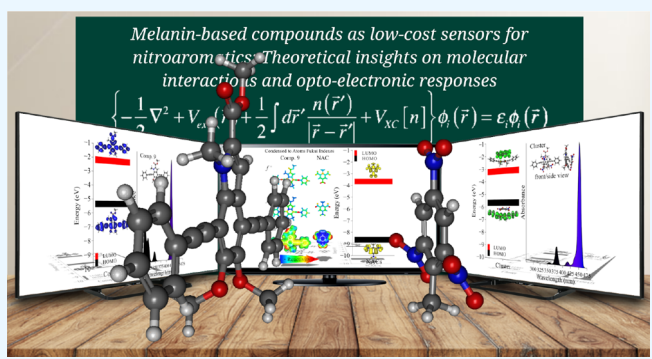

**ABSTRACT:** Nitroaromatic compounds (NACs) are used in various industrial applications including dyes, inks, herbicides, pharmaceuticals, and explosives. Due to their toxicity and environmental persistence, reliable detection and monitoring methods are required. Hybrid organic−inorganic structures have shown potential for NAC sensing; however, their complex synthesis, high processing costs, and limited reproducibility hinder practical implementation, highlighting the need for simpler and more accessible materials. In this study, we employed density functional theory (DFT)-based calculations to evaluate the electronic, optical, and reactive properties of two melanin-based oligomeric systems, aiming to assess their potential use as NAC detectors. Our results indicate the potential of these materials to detect a series of nitroaromatic compounds such as 2,4-DNP, 2,4-DNT, 2,6-DNT, TNP, and TNT by electrical and infrared optical measurements. Born−Oppenheimer molecular dynamics (BOMD) simulations reveal the thermal stability of the adsorption process, confirming effective substrate−analyte interaction under different temperature conditions. To the best of our knowledge, this compound has not been proposed for sensing applications. Its low cost and facile synthesis make it a promising candidate for the development of environmentally friendly organic NAC sensors.

## 1. INTRODUCTION

Nitroaromatic compounds (NACs) are aromatic structures with one or more nitro groups ($-NO_2$). The presence of the $-NO_2$ group makes NACs useful as raw materials in the chemical syntheses of a variety of compounds such as corrosion inhibitors, antioxidants, preservatives, fuel additives, dyes, paints, cosmetics, fungicides, herbicides, pesticides, drugs, and other industrial chemicals.[1−3] NACs are of primary concern as they are mutagenic and carcinogenic,[4] as well as toxic to living organisms.[2,3] Nitro groups make NACs recalcitrant; therefore, their degradation is not sustainable and effective, leading to their accumulation in the environment and making NACs a serious threat to the ecological environment and human health.[3]

NACs, such as nitrobenzenes (NB), can cause diseases such as anemia, skin irritation, and cancer.[5] NB poisoning in humans causes methemoglobin formation, cyanosis, neurotoxic effects, unconsciousness, gastric irritation, nausea, vomiting, drowsiness, convulsions, coma, respiratory failure, and may result in death.[6−8] In addition, NB can be metabolized to *p*-aminophenol and *p*-nitrophenol, being very slowly eliminated by the organism.[9]

The development of materials and devices for detecting NACs is therefore essential. It has seen a resurgence since the 2000s in particular because NACs were used as explosives in some terrorist attacks,[10,11] giving rise to several detectors.[12−14] In particular pyridine, diazine, and triazine have been studied in detail due to their properties and their use as chemical sensors for chemical analyses.[15,16] In recent years, other types of sensors have been proposed such as the Mach−Zehnder interferometer waveguide sensor using porous polycarbonate, with fast responses and high sensitivity. Optical sensors have also been proposed via the Förster resonance energy transfer (FRET) mechanism,[17] PbS quantum dots,[18−20] and hybrid perovskites.[21] Metal−organic complexes, such as MOFs (metal−organic frameworks) and rare-earth metal-based luminescent coordination polymers (LCPs), have also been considered for NAC detection, mainly due to their tunable porosity, optical properties, and analyte affinity.[15,22−24] Although these compounds show promising sensing performance, their practical application is hindered by synthetic complexity and processing challenges. In particular, complex crystal engineering, multistep routes, and occasional reliance on unexpected transformations have been reported by Dutta et al.[25] Some energetic MOFs require costly components,[26] presenting low hydrothermal and chemical stabilities. Difficulties in relation to regeneration and recycling have also been reported, which further complicates their practical use.[23,27] Although some specific MOFs present scalable and low-cost production, their crystals are inherently brittle in nature and arduous to process for practical applications.[26]

Some of the disadvantages identified above could potentially be mitigated by using organic-based materials as sensors. Specifically, melanins have shown promise in various applications, including pH sensors,[28,29] relative humidity sensors,[30] solar cells,[31,32] and organic light-emitting diodes (OLEDs).[33] However, the use of such materials for NAC detection remains largely unexplored. The difficulties associated with the structural characteristics of natural melanins and the resulting lack of reproducibility of the experiments have led to the use of synthetic melanin derivatives for the active layer of these devices. Understanding the complex physical and chemical properties of such melanin-based materials has broadened the prospect of their application in devices,[34,35] prompting us to investigate the possibility of their use in sensors.

In particular, Selvaraju et al. have proposed a series of molecules with melanin-inspired cores for optoelectronic applications.[36,37] These compounds are synthetically accessible in good yields from renewable precursors (e.g., vanillin), and they exhibit compatibility with standard cross-coupling methodologies. They exhibit high solubility and display photophysical and electrochemical properties suitable for stable integration into optoelectronic devices.[32] In addition to the melanin-based core, these derivatives possess electron-rich C≡C bonds that facilitate conjugation and delocalization, while reinforcing molecular rigidity and planarity that are essential for efficient charge transport in organic materials,[38−40] as well as charge transfer and molecular recognition in sensing platforms.[38] Moreover, these structures are functionalized with electron-donating methoxy (−OCH$_3$) groups, which act as strong electron donors,[41] increasing the electron density of the aromatic ring and favoring interactions with electron-deficient analytes (such as NACs). Compared to other electron-donating groups (e.g., −OC$_2$H$_5$), methoxy offers a favorable combination of electronic enhancement and low steric hindrance, helping preserve the planarity and π-conjugation of the backbone, relevant for charge transfer and sensitivity.[41] Previous studies have shown that methoxy substitution can modulate electronic properties (reducing the HOMO−LUMO gap) and enhance the optoelectronic performance of conjugated systems,[42] supporting its role in the design of functional sensing materials.

These insights motivate the use of computational modeling to further investigate the sensing potential of such melanin derivatives and to guide future experimental efforts toward the development of new compounds with improved performance. Given the limitations of many experimental approaches in resolving molecular-level interactions, computational modeling has become a powerful and cost-effective strategy for predicting sensor performance, estimating binding affinities,[43] and guiding the rational design of sensing materials.[44] In this context, theoretical investigations were employed to evaluate the potential of melanin-inspired compounds 9a and 9b, reported by Selvaraju et al.,[36] as NAC detectors. Electronic structure calculations and molecular dynamics were performed for such monomeric structures, and the effects of a variety of nitroaromatics were evaluated by using density functional theory (DFT)-based calculations. The results indicate that melanin-inspired compounds 9a and 9b exhibit strong and thermally stable interactions with nitroaromatics (notably TNT and TNP), inducing measurable electronic and vibrational shifts. These findings position melanin-inspired compounds as promising, low-cost materials for NAC sensing.

## 2. MATERIAL AND METHODS

**2.1. Materials.** Figure 1 shows the structures that were considered in this study. For simplicity, the compound denomination used in ref 36 was kept (9a and 9b, see Figure 1a,b). Figure 1c shows the NACs that were considered as analytes: nitrobenzene (NB), o-nitrophenol (o-NP), m-nitrophenol (m-NP), p-nitrophenol (p-NP), o-nitrotoluene (o-NT), m-nitrotoluene (m-NT), p-nitrotoluene (p-NT), 1,3-dinitrobenzene (1,3-DNB), 2,4-dinitrophenol (2,4-DNP), 2,4-dinitrotoluene (2,4-DNT), 2,6-dinitrotoluene (2,6-DNT), trinitrophenol (TNP), and trinitrotoluene (TNT).

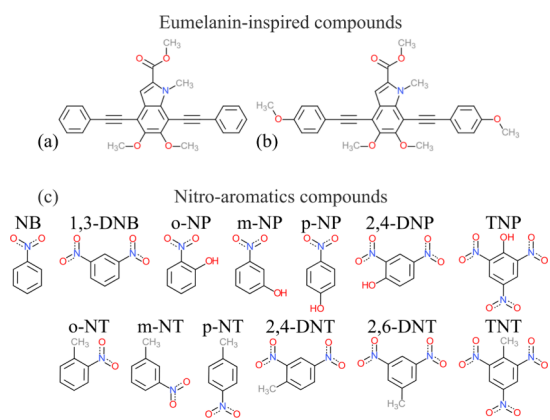

**Figure 1.** Chemical structures of melanin-inspired compounds 9a (a) and 9b (b) (substrates). Chemical structures of NACs (c) (analytes).

**2.2. Methodology.** The structures were designed with the aid of the GaussView computational package.[45] Conformational searches were conducted via molecular dynamics (MD) simulations at high temperatures (Amber Potential at 1000 K of temperature with the aid of Gabedit software[46]). The lowest energy conformer (coming from MD) was fully optimized in the framework of density functional theory (DFT) using the B3LYP[47,48] exchange-correlation (XC) functional and the 6−311G(d,p) basis set on all the atoms.

Local reactivities were evaluated via the condensed-to-atoms Fukui indexes (CAFIs),[49,50] molecular electrostatic potentials (MEPs),[51] and the spatial distribution of the frontier molecular orbitals (FMOs, i.e., the highest occupied and the lowest unoccupied molecular orbitals, HOMO and LUMO, respectively).

The relative alignments between the FMO energies of the melanin-based oligomers and the NACs were evaluated to assess the applicability of these systems as chemical sensors, taking into account the possible effects of the analytes on the substrate. Figure 2 illustrates some possible effects of analytes

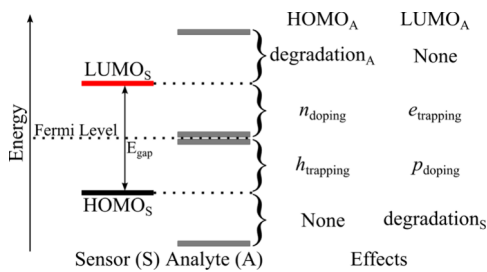

**Figure 2.** Relative alignments between the FMO energies of the sensor active layer (S) and the analytes (A) and possible electrical effects.

(A) on the sensor (S) electrical response expected for distinct FMOs relative alignments. Red and black lines represent the FMOs of the sensor, and gray ones are those of the analyte (S and A subscripts are used for simplicity, respectively). The diagram shows the possible effects according to $HOMO_A$ and $LUMO_A$ relative positions: (i) material degradation due to charge transfer processes for $HOMO_A > LUMO_S$ and $LUMO_A < HOMO_S$ (analyte and sensor degradation, respectively); (ii) nonappreciable electric responses are expected for the configurations where $LUMO_A > LUMO_S$ and $HOMO_A < HOMO_S$, once occupied and unoccupied levels of the A are inserted, respectively, in the valence and conduction bands of S; (iii) electrochemical doping and charge trapping are expected when the FMOs of A are inserted into the band gap of S, depending on their relative positions in relation to the Fermi level of S ($FL_S$), e.g., n-doping is expected when $FL < HOMO_A < LUMO_S$ while hole trapping ($h_{trapping}$) effects are expected when $HOMO_S < HOMO_A < FL_S$; similarly, we have p-doping for $HOMO_S < LUMO_A < FL_S$ and electron trapping ($e_{trapping}$) for $FL_S < LUMO_A < LUMO_S$.[44,52]

The HOMO and LUMO energies ($E_{HOMO}$ and $E_{LUMO}$) of all systems were estimated via Kohn–Sham eigenvalues (KS) and compared with those reported elsewhere.[53–55] The electronic gaps were estimated by $E_{gap} = E_{LUMO} - E_{HOMO}$. The optical properties of 9a and 9b (in particular the optical gap, $E_{opt}$) were estimated via time-dependent (TD) DFT calculations, by using the same functional and basis set (i.e., TD-DFT/B3LYP/6−311G(d,p) approach).

The donation and acceptance indexes ($R_D/R_A$) were estimated from the analysis of the relative electron-accepting ($\omega^+$) and electron-donating ($\omega^-$) powers of the compounds, estimated by[56,57]

$$\omega^- = \frac{(3IP + EA)^2}{16(IP - EA)} \quad (1)$$

$$\omega^+ = \frac{(IP + 3EA)^2}{16(IP - EA)} \quad (2)$$

where $IP = E(N-1) - E(N)$ and $EA = E(N) - E(N+1)$ represent, respectively, the ionization potential and electron affinity of the molecules. The $R_D$ and $R_A$ indexes are obtained by comparing $\omega^+$ and $\omega^-$ powers with those of sodium ($\omega^-_{Na}$ = 3.46) and fluorine ($\omega^+_F$ = 3.40), respectively:[56,57]

$$R_D = \frac{\omega^-}{\omega^-_{Na}} \quad (3)$$

$$R_A = \frac{\omega^+}{\omega^+_F} \quad (4)$$

which are associated with the charge transfer capacity of the compounds. All the calculations were conducted with the aid of the Gaussian 16 computational package.[58]

The analytes that exhibited greater potential for detection by melanin-based compounds were considered in the adsorption studies. For this purpose, two distinct procedures were considered to generate substrate + analyte clusters:

1. *Adsorption guided by CAFIs:* the analytes were manually placed over the substrate structures considering the alignment of high CAFI values (e.g., the analytes were positioned so that their most reactive sites were close to the triple bonds of the melanin compound with a distance of 1.5 Å) and subjected to geometry optimization in a DFT/B3LYP/6−311G(d,p)/GD3 approach,

2. *Adsorption via docking submodule by automated interaction site screening (aISS):*[59] done via the aISS package and subjected to tight-binding geometry optimization (GFN2-xTB) to select more stable structures, which were further optimized in the DFT/B3LYP/6−311G-(d,p)/GD3 approach,

These systems were subjected to full geometry optimization and interaction calculations. All the calculations for the adsorbed systems were conducted considering the D3 version of Grimme's dispersion correction (GD3).[60] The complexation energies were estimated using the counterpoise method to correct the basis set superposition error (BSSE).[61,62] The evaluation of partial density of states (PDOS) and weak interactions[63,64] was conducted with the aid of the MultiWFN computational package.[65]

Adsorbed structures stabilities were evaluated via NVT Born–Oppenheimer molecular dynamics (BOMD) simulations for selected systems (isolated compounds and those adsorbed with TNT and TNP) with the aid of DFTB+ software within the DFTB3 formalism.[66,67] Distinct temperatures were considered in the simulations (300, 400, 500, and 650 K), using a Nosé–Hoover thermostat. The Slater–Koster parameters were selected from the "3ob-1−1" set due to their excellent agreement with simulations conducted using the B3LYP functional,[68] ensuring consistency with the DFT approach methodology. DFT-D3 dispersion corrections were also incorporated.[69] A self-consistent charge (SCC) tolerance of $10^{-6}$ over a total simulation time of 100 ps was considered with a time step of 0.97 fs (∼10 times the period associated with the highest vibrational frequency of each configuration).

The stability of each adsorbed system was evaluated from the time-averaged density distribution, $\rho(r_\pi)$, of the distance between the centers of mass of the analyte and the substrate, $r_\pi$, as illustrated in Figure 3.

Furthermore, to gain insight into the vibrational analysis, the autocorrelation approach for atomic velocities[70–72] and dipole moments was employed.[73,74] Given an intensive property, such as an atom's velocity, $\vec{v}_i(t)$, or the system's dipole moment,

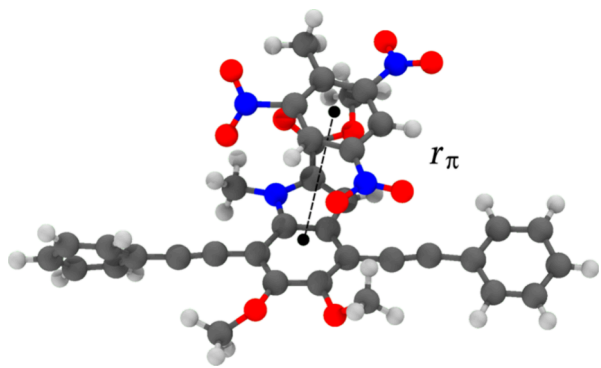

**Figure 3.** Illustration of $r_\pi$, the distance between the centers of mass (depicted as black dots), for the 9a compound with the TNT analyte. This distance is tracked throughout each BOMD trajectory to obtain the $\rho(r_\pi)$.

$\vec{\mu}(t)$, the normalized autocorrelation function can be computed, as expressed in eq 5.

$$C_v(t) = \langle \vec{v}_i(0) \cdot \vec{v}_i(t) \rangle \quad (5)$$

where $\vec{v}_i(t)$ and $\vec{\mu}(t)$ represent the intensive properties of interest at time $t$. These trajectories were sampled within 1 ps windows and averaged over a total trajectory time of 100 ps. As established in the literature, the Fourier transform of velocity and dipole moment autocorrelations provides insight into the vibrational density of states (VDOS) and infrared (IR) spectra.[75] The peaks obtained can reveal infrared absorption properties, displaying vibrational signatures typically usually observed in first-order Raman and IR experimental spectra.

## 3. RESULTS AND DISCUSSION

**3.1. Isolated Structures.** Table 1 summarizes the optoelectronic properties of compounds 9a and 9b, as well as experimental values reported in ref 36, estimated from the onset of the first oxidation and reduction potentials (in parentheses). As can be seen, the theoretical results present a reasonable agreement with the experimental values, mainly regarding the optical band gaps. The theoretical evaluation of the oligomers' optical properties makes a correlatable, self-consistent estimation of the optical behavior of the 9a and 9b systems. Table 2 summarizes the electronic properties of NACs, which are in agreement with the values reported in the literature, measured by cyclic voltammetry, XPS, and estimated by density functional theory.[17,53−55]

To first estimate the applicability of melanin-inspired 9a and 9b compounds as NAC sensors, comparative analyses of the relative alignments between their FMOs and the distinct analytes were conducted (Figure 4). The dashed lines in Figure 4 indicate the position of the FMOs, and the dotted line represents the Fermi Level of the nondoped systems ($E_F = E_g/2$).

As a matter of fact, several factors can influence the efficiency of organic sensor devices. An important aspect is the relative position of the FMOs of the analytes in relation to the electronic gap of the active compounds.[52] From Figure 4, it is noticed that the 9a and 9b monomers appear to be promising structures for NAC detection, mainly in relation to di- and trinitroaromatics. It should be noted that TNT can act as a *p*-type dopant for compound 9b, while TNP can act as a *p*-type dopant for both structures 9a and 9b. In general, the relative positions of the FMOs allow us to suppose that NACs should act as electron traps in 9a or 9b, and then influence the optoelectronic properties of these materials.

**Table 2. Summary of Theoretical Electronic Properties of the NACs**

| compound | abbreviation | $E_{HOMO}$ (eV) this study (literature) | $E_{LUMO}$ (eV) this study (literature) |
|---|---|---|---|
| nitrobenzene | NB | −7.82 (−7.59) | −2.63 (−2.43) |
| 1,3-dinitrobenzene | 1,3-DNB | −8.62 (−8.41) | −3.32 (−3.14) |
| *ortho*-nitrophenol | *o*-NP | −7.04 (−7.21) | −2.32 (−2.23) |
| *meta*-nitrophenol | *m*-NP | −7.01 (−7.18) | −2.59 (−2.88) |
| *para*-nitrophenol | *p*-NP | −7.14 (−7.35) | −2.42 (−2.98) |
| 2,4-dinitrophenol | 2,4-DNP | −7.88 (−7.63) | −3.01 (−3.32) |
| trinitrophenol | TNP | −8.41 (−8.24) | −4.05 (−3.90) |
| *ortho*-nitrotoluene | *o*-NT | −7.50 (−7.28) | −2.53 (−2.31) |
| *meta*-nitrotoluene | *m*-NT | −7.48 (−7.27) | −2.56 (−2.36) |
| *para*-nitrotoluene | *p*-NT | −7.57 (−7.57) | −2.51 (−2.50) |
| 2,4-dinitrotoluene | 2,4-DNT | −8.31 (−8.11) | −3.16 (−2.98) |
| 2,6-dinitrotoluene | 2,6-DNT | −8.10 (−7.27) | −3.03 (−2.36) |
| trinitrotoluene | TNT | −8.65 (−8.46) | −3.65 (−3.50) |

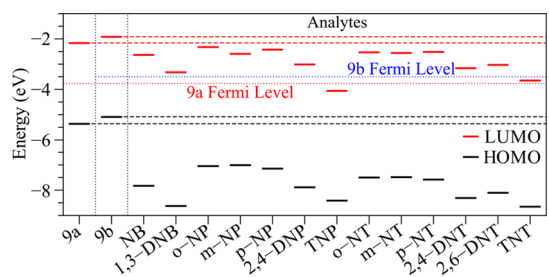

**Figure 4.** Comparative analyses of the FMO relative alignments of melanin-inspired compounds in relation to nitroaromatics.

In particular, the results presented in Figure 4 suggest that the presence of NACs can induce significant changes in electron transport mechanisms (and also in charge recombination) that could be monitored in electron-only devices (or ambipolar devices) via electrical (or optical) characterization (e.g., changes in current densities, electrical impedance, absorbance, and so forth).

To better interpret possible charge transfer effects between the structures, the donor−acceptor electron map (DAM) is presented in Figure 5. This map allows us to classify the systems as electron-donating ($R_d$) and electron-accepting ($R_a$) compounds. In general, low $R_d$ values indicate good donors, while high $R_a$ values define good acceptors (as indicated by the red arrows).

**Table 1. Summary of Optoelectronic Properties of 9a and 9b Eumelanin-Based Compounds**

| compound | method | $E_{HOMO}$ (eV) | $E_{LUMO}$ (eV) | $E_{gap}$ (eV) | $E_{opt}$ (eV) |
|---|---|---|---|---|---|
| 9a | theory (Exp.) | −5.36 (−5.55) | −2.16 (−2.70) | 3.19 (2.85) | 2.94 (2.94) |
| 9b | theory (Exp.) | −5.08 (−5.45) | −1.92 (−2.65) | 3.17 (2.80) | 2.90 (2.87) |

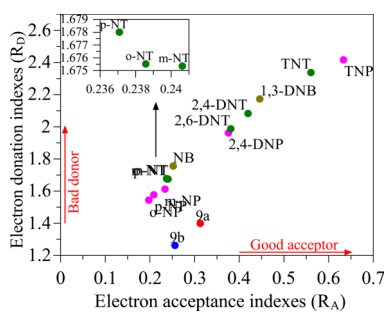

Figure 5. Comparative analyses of the electron donation and acceptance indexes of melanin-inspired oligomers and NACs.

It can be seen that the 9a and 9b monomeric structures are better donors than the NACs (the differences between 9a and 9b are due to the terminal methoxy groups). Trinitroaromatics, in particular, are good electron acceptors and poor donors, followed by di- and mononitroaromatics. In particular, the higher electron affinity of TNT and TNP indicates an effective interaction of these analytes with the monomers 9a and 9b.

From Figures 3, 4, and S1, a stronger interaction of the monomers with the NACs 1,3-DNB, 2,4-DNP, 2,4-DNT, 2,6-DNT, TNP, and TNT can be deduced, considering their ability to insert unoccupied states into the 9a and 9b band gaps and their corresponding electron acceptor/donation indices. For this reason, only these analytes were selected for the adsorption studies.

To interpret the interaction between compound 9 and NACs, the local reactivity of the compounds was investigated. Figures 6 and 7 summarize the CAFIs and MEPs of the NACs and the structures of the compounds. Red and blue sites presented in the CAFI (MEP) maps represent, respectively, reactive (negatively charged) and nonreactive (positively charged) sites. In general, sites with higher values of $f^+$, $f^-$, and $f^0$ (red sites) represent regions that are prone to interact with nucleophiles (being prone to receive electrons), electrophiles (losing electrons), and free radicals (with no changes in the total number of electrons), respectively.

It should be noted that electron acceptance of nitroaromatic compounds is concentrated on the nitro groups (i.e., high $f^+$ values), while electron donation is centered on the ring atoms for compounds with one nitro group and on -$NO_2$ for compounds with two or three nitro groups (i.e., high $f^-$ values). Hydroxyl groups also play an important role in relation to $f^-$. The most reactive regions of compounds 9a and 9b are centered on the C≡C groups in both structures, suggesting that these regions are the most important sites for charge transfer processes.

### 3.2. Adsorbed Structures.

All the adsorbed structures obtained by the docking submodule (aSSI) exhibited higher energy values compared to the structures from CAFI's guided adsorption method after geometry optimization, even those structures that showed hydrogen bonds are less energetic (see Figure S2). Such results evidence the relevance of considering CAFIs as effective adsorption center predictors, as already proposed elsewhere.[44,76,77] In this sense, for simplicity, only the results coming from CAFI-based methods are presented (results coming from aSSI are shown in the Supporting Information).

Figure 8 shows the spatial and energy distributions and Kohn−Sham frontier molecular orbitals of the adsorbed systems.

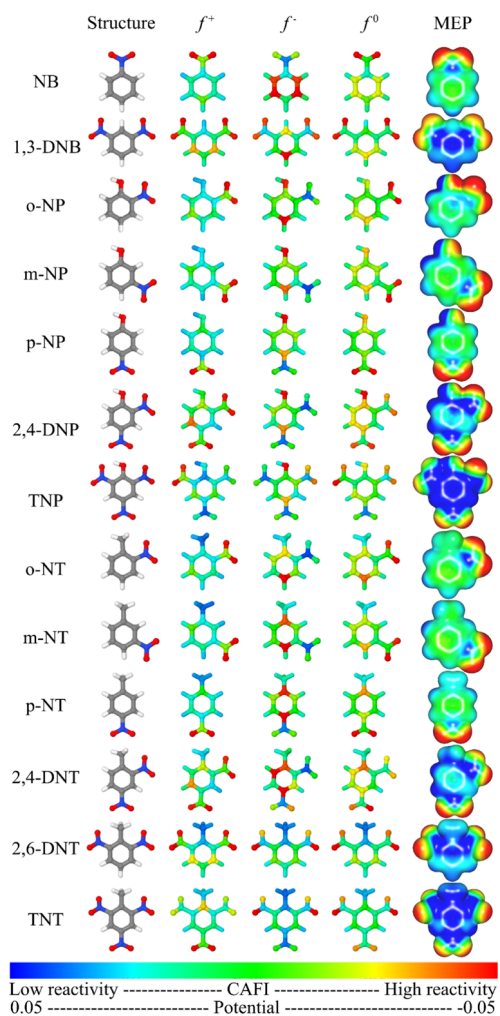

Figure 6. CAFIs and MEPs estimated for NACs.

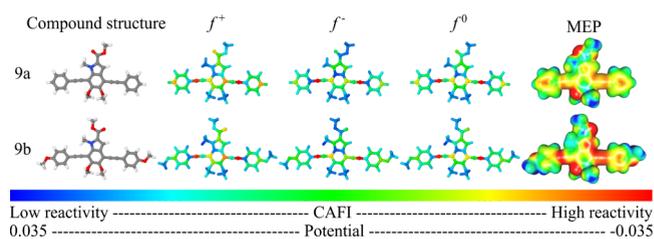

Figure 7. CAFIs and MEPs of monomeric structures of melanin-inspired oligomers.

It should be noted that in both cases, the HOMO is localized on the melanin-based compound, while the LUMO is mainly located on the analytes. As preliminarily predicted in Figure 5 and confirmed by CAFI (Figures S3 and S4 in the Supporting Information), the LUMO energy level of the adsorbed structure is primarily influenced by the analytes, resulting in a smaller band gap compared to the isolated compound.

Figures 9 and 10 illustrate the partial and total density of states (PDOS and DOS) representations of the adsorbed structures that evidence the dominance of the melanin-based substrates and analytes on the HOMO and LUMO, respectively. Red, blue, and green curves define the PDOS of compound 9a, compound 9b, and the analytes, respectively.

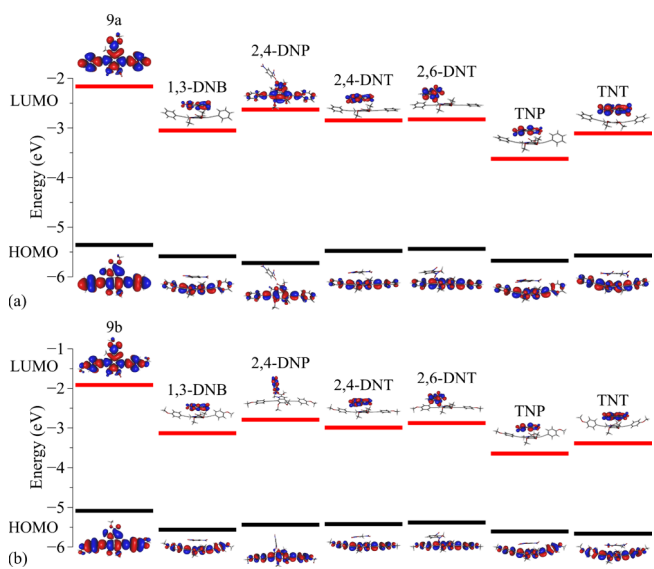

Figure 8. Spatial distribution and energy levels of the FMOs over the monomer and analytes: (a) compound 9a and (b) compound 9b.

The position of the HOMO is indicated by the vertical dashed line.

Similarly to Figure 8, all of the HOMOs (dashed lines) are dominated by the melanin-based compound, while the LUMO is predominantly associated with the analytes (similar results are shown in Figures S8 and S9). It should be noted that FMO alignments evidence an effective electron trapping behavior of the analytes, with potential implications in photoluminescence and exciton dynamics, by photoinduced electron transfer. Indeed, a number of studies have reported the effective fluorescence quenching induced by nitroaromatics (specially TNT).[17,78] A similar effect should take place for 9a and 9b, once they present high photoluminescence quantum yields.[36] In particular, higher spatial overlap matrix elements $\langle |\varphi_{HOMO}\| \varphi_{LUMO}|\rangle$ (which play a key role in fluorescence quenching) are observed for 1,3-DNB, TNP, and TNT, suggesting enhanced sensitivity to these compounds (see Supporting Information).

To better evaluate the compound + analyte interaction, the complexation energies (Figure 11) and weak interaction areas (Figure 12) were investigated. Complexation energies are widely used as essential descriptors of sensor performance. In general, absolute values lower than 0.5 eV indicate weak physisorption, while those in the range of 0.6−1.2 eV are considered optimal, offering a balance between binding strength and desorption efficiency. Absolute values exceeding 1.2 eV typically reflect strong chemisorption, which may hinder analyte desorption and sensor reusability.[79−83]

Figure 11 reveals lower complexation energies in both systems (9a and 9b) when interacting with 2,4-DNP and 2,6-DNT, which is consistent with the smaller interaction area presented in Figure 12. On the other hand, higher complexation energies and interaction areas are observed with TNP and TNT. It should be noted that absolute values around 0.6−1.0 eV are obtained for all the systems, combining adequate binding with reversible analyte release. In particular, our melanin-inspired systems exhibit interaction strengths com-parable to those of established materials, including $C_5N_2$ (−1.37 to −1.49 eV for TNT and PA)[84] and Pd-decorated $MoSi_2N_4$ (−1.21 eV for nitrobenzene).[85]

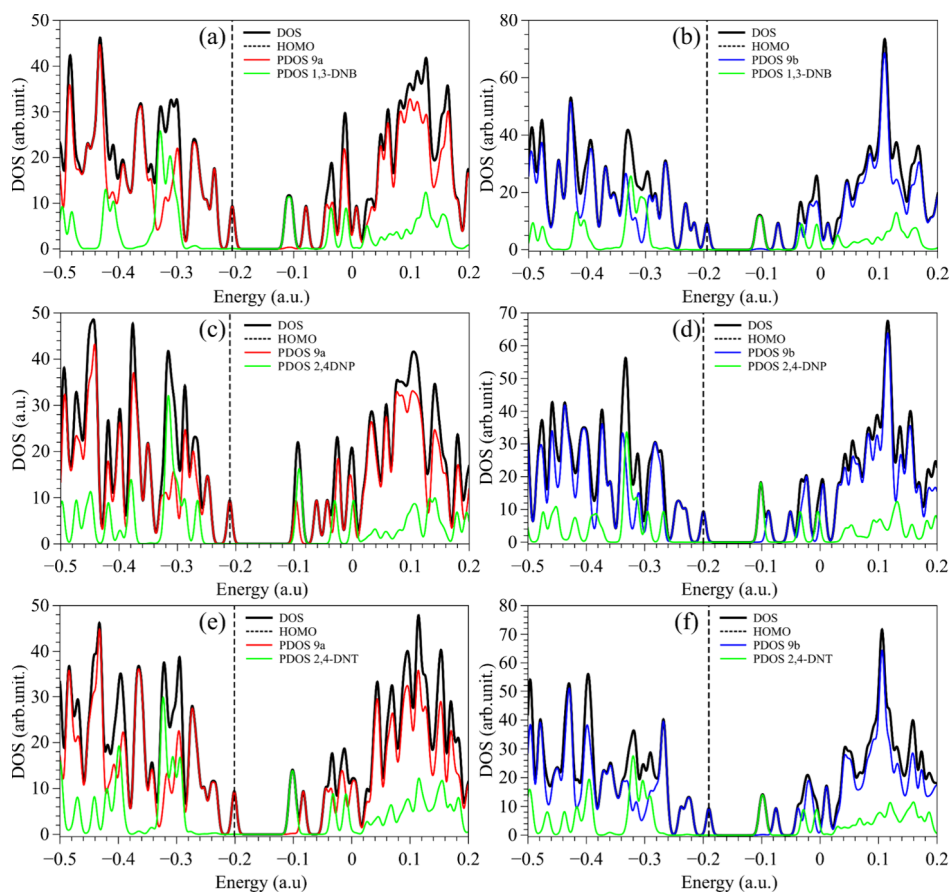

Figure 9. DOS and PDOS of melanin-inspired compounds 9a (left) and 9b (right) with (a, b) 1,3-DNB, (c, d) 2,4-DNP, and (e, f) 2,4-DNT.

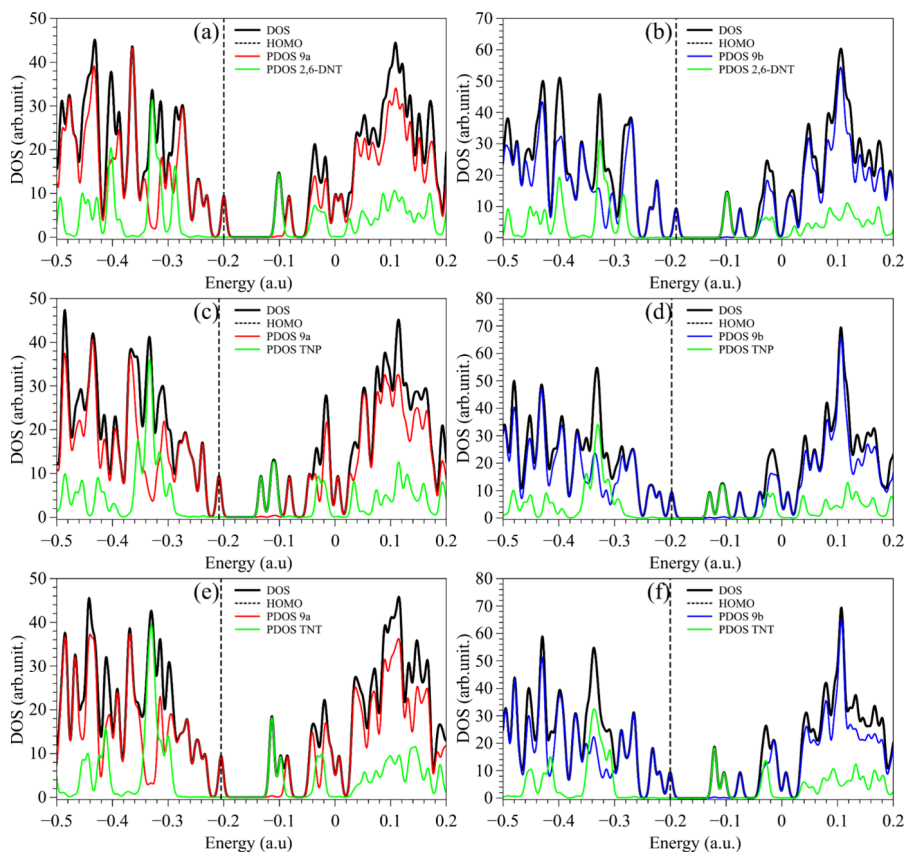

**Figure 10.** DOS and PDOS of melanin-inspired compounds 9a (left) and 9b (right) with (a, b) 2,6-DNT, (c, d) TNP, and (e, f) TNT.

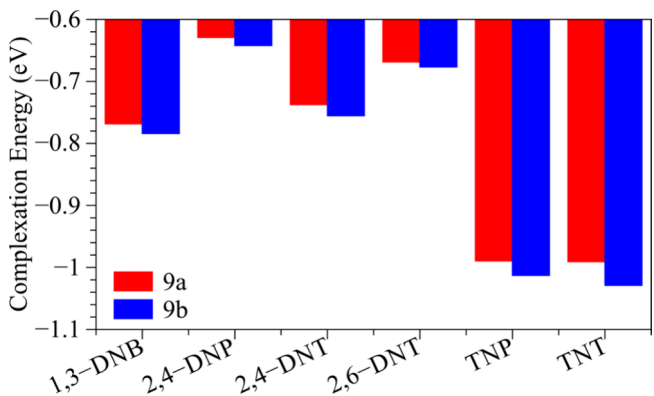

**Figure 11.** Complexation energy of analytes on melanin-inspired compounds.

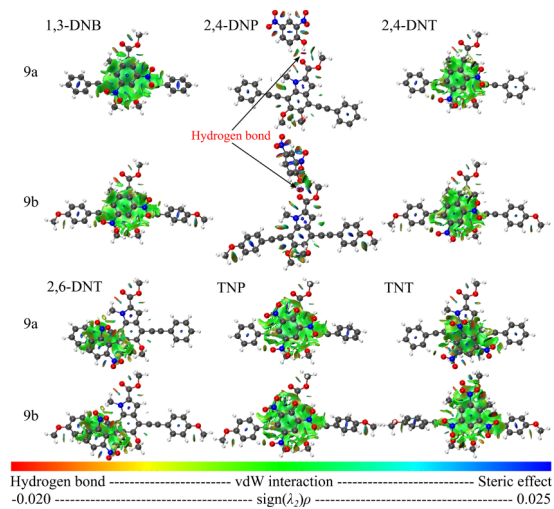

**Figure 12.** Analyte-melanin-based compound interactions: strength and interaction areas.

It is worth noting that most systems exhibit significant van der Waals (vdW) interactions (highlighted in green and yellow), with only the 2,4-DNP complex displaying a hydrogen bond. This specific interaction arises from the particular geometry adopted during optimization and influences the nature of the electronic transitions: in the 9a + 2,4-DNP complex, both the HOMO and LUMO are localized on the substrate (compound 9a), whereas in the 9b + 2,4-DNP complex, the HOMO is localized on the substrate and the LUMO on the analyte (Figures 8 and 9). Interestingly, systems obtained via the automated docking method (aISS) adopted similar configurations to those guided by CAFI (see Figure S10, Supporting Information), displaying dominant π−π stacking interactions. These structures consistently showed HOMO localization on the substrate and LUMO on the analyte, along with higher complexation energies, highlighting a preferred orientation for charge transfer processes.

Regardless of the adsorption method employed, both approaches indicated strong and favorable interactions between the melanin-based compounds and analytes such as TNT and TNP, reinforcing the robustness of the interaction pattern across computational protocols.

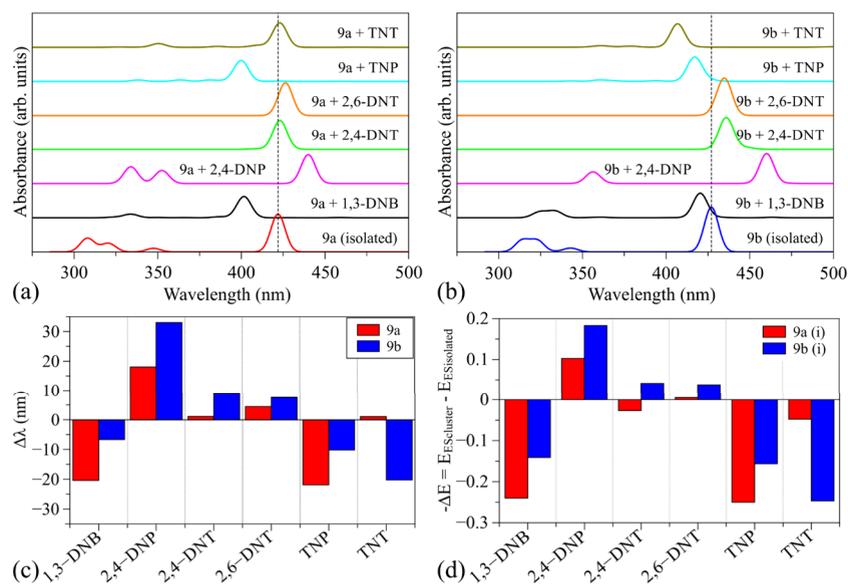

**Figure 13.** Theoretical optical absorption spectra of (a) compounds 9a and (b) 9b: isolated and adsorbed with NACs (Gaussian curves with a half width of 5 nm). The variations in (c) wavelength absorption and (d) excited-state energy of the cluster compared to the isolated structure.

To investigate the selectivity of the compounds, additional adsorption studies were carried out with common atmospheric compounds, $N_2$ and $O_2$ (at the triplet state), using the same theoretical approach used for the NACs. Both analytes exhibited low adsorption: the interaction energies for $N_2$ ($O_2$) were around 8x (10x) lower than those calculated for TNP and TNT. These results suggest noneffective interactions, evidencing the selectivity of our systems toward NACs. All corresponding energy values are detailed in Table S3 of the Supporting Information.

Additional information regarding 9a/b + NACs system stability was assessed by recovery time ($\tau$) estimation (time required for analyte desorption from the substrate[85]), which shows $\tau$ ranging from a few hours for T = 300 K up to microseconds for T = 650 K under visible light irradiation (see Table S2 in the Supporting Information for details).

To estimate the possible optical response of such a melanin-based substrate to the analytes, additional calculations were conducted for the absorbed systems in the framework of the TD-DFT/B3LYP/6−311G(d,p). Figure 13a,b depicts the absorption spectra of compounds 9a and 9b isolated and adsorbed with distinct NACs, as well as the main peak shift noticed for each substrate/analyte system; Figure 13c presents the numerical shift observed in Figure 13a,b. Figure 13d shows the negative variation of the excited-state energy for the most representative transition in the vertical transition.

It should be noted that, in general, NAC adsorption leads to significant changes in the main peak optical absorption of the substrates, which depends on compound 9a or 9b. Some interesting trends can be observed, dividing the compounds in hypsochromic (blue-shifted: 1,3-DNB; and TNP), bathochromic (red-shifted: 2,4-DNP; 2,4-DNT; and 2,6-DNT), and anomalous (with no pattern: TNT) analytes.

Significant deviations are noticed for 1,3-DNB ($\Delta\lambda$ = −20.4 nm), 2,4-DNP ($\Delta\lambda$ = +18.1 nm), and TNP ($\Delta\lambda$ = −21.9 nm) in relation to compound 9a, with very small changes for the others (i.e., $\Delta\lambda$ < 5 nm). For compound 9b, the most relevant optical responses were observed for 2,4-DNP ($\Delta\lambda$ = 33.0 nm) and TNT ($\Delta\lambda$ = −20.2 nm), with intermediate responses for the other analytes: 1,3-DNB ($\Delta\lambda$ = −6.6 nm); 2,4-DNT ($\Delta\lambda$ = 8.9 nm); 2,6-DNT ($\Delta\lambda$ = 7.6 nm); and TNP ($\Delta\lambda$ = −10.2 nm).

These changes can be rationalized in terms of inductive effects and/or the introduction of new electronic states within the substrate band gaps.[77] In fact, the insertion of empty levels (analytes' LUMOs) inside the 9a and 9b gaps should lead to systems with reduced band gaps, as indeed observed in Figures 8−10. Such changes were supposed to result in bathochromic optical effects for all the systems, with a relative amplitude of 96 ± 49 nm (for 9a) and 135 ± 57 nm (for 9b) (compatible with $\Delta E_{gap}$ −0.6 ± 0.2 and −0.8 ± 0.2 eV, respectively), which is indeed observed with very small amplitude ($<1.4 \times 10^{-2}$) for systems 9a + 1,3-DNB and 9b + 1,3-DNB respectively.

The distinct dominant optical responses obtained for the systems are associated with the low superposition of the resulting FMOs, as evidenced in Figure 8, indicating a low probability of HOMO (old—9a/9b centered) to LUMO (new—analyte centered) transitions and showing mainly H-L$_2$ or H-L$_3$ (9a/9b centered) transitions (see Figures 9 and 10, as well as Table S1 in the SI). Figure 13c shows the variation of $E_{ES}$ (excited-state energy) of the cluster in relation to the isolated compounds. It is important to note that greater variations in the energies of excited states lead to larger shifts in the optical absorption spectrum. With the exception of 9a with 2,4-DNT and TNT clusters, a decrease in energy results in a red shift, while an increase in energy results in a blue shift.

The resulting spectra are governed by inductive effects and small perturbations of the electronic structures in the presence of intermediate levels. In particular it is noticed that effective interactions between reactive oxygen atoms of the nitro groups (with high $f^+$ values) of NACs with substrate triple bonds (with high $f^-$ values) lead to significant hypsochromic effects noticed for 1,3-DNB, TNP (for compounds 9a and 9b), and TNT (for compound 9b). This configuration indicates an effective substrate-to-analyte electron transfer process, which weakens the $\pi$-systems of the substrates, reducing their effective conjugation lengths and promoting the hypsochromic responses. The absence of significant changes on the 9a + TNT system in relation to 9b + TNT is due to the absence of $NO_2$-triple bond interaction noticed for 9b (replaced by $CH_3$- triple bond interaction). The redshift associated with 2,4-DNP is linked to the formation of O−H bonds, which improves the aromaticity on the central rings of the substrates. The observed variability in optical absorption shifts may be attributed to the diversity of interaction types (π−π stacking, NO$_2$−C≡C interactions, and hydrogen bonding) and the specific adsorption geometries adopted by each analyte. While such orientation differences influence local electronic transitions and complexation energies, they do not significantly alter the overall HOMO−LUMO gap closure, which remains consistently reduced across systems. The combined analysis of adsorption energies and frontier molecular orbitals (including relative alignments and spatial overlaps) provides a useful metric for evaluating the sensor's relative sensitivity to each analyte. In particular, the higher spatial overlap matrix elements and stronger adsorption energies observed for TNP and TNT support their selection for further stability assessment via BOMD simulations.

These theoretical findings can be meaningfully compared with experimental data from similar eumelanin-inspired molecules.[37] Notably, Selvaraju et al. reported that indole-based conjugated systems with phenylene ethynylene linkers exhibit modulated HOMO−LUMO energy levels and band gaps depending on terminal substituent behavior that parallels the analyte-induced bandgap shifts observed in our work. Importantly, their study shows that nitroaromatics effectively quench photoluminescence, attributed to LUMO localization on the NO$_2$-containing analyte and HOMO retention on the substrate, thus facilitating photoinduced electron transfer (PET). This agrees with the orbital alignments and spatial overlaps observed in our adsorbed systems, particularly for TNP and TNT. The consistent HOMO−LUMO separation and electronic coupling strongly support fluorescence quenching as a more robust sensing mechanism. These insights highlight the importance of future experimental studies of photoluminescent responses for validating and expanding the detection capabilities of melanin-inspired platforms.

### 3.3. Born−Oppenheimer Molecular Dynamics.

Figures 14 and 15 summarize key results derived from the Born−Oppenheimer molecular dynamics (BOMD) simulations, providing dynamic insights into the structural stability and vibrational behavior of the analyte−substrate complexes under thermal stress.

Figure 14 shows the time-averaged density distribution of the distance between the analyte/substrate centers of mass, coming from BOMD simulations. It should be noted that across all systems, increasing the temperature (and consequently the kinetic energy) leads to a greater average displacement of the analyte from its initial position, reflected in broader $\rho(r_\pi)$ distributions and decreased peak intensity. This behavior is consistent with reduced interaction strength and higher desorption probabilities at elevated temperatures.

Notably, although the time scales explored in the simulations are shorter than those expected for analyte dissociation at ambient temperature and 400 K, the broadening of $\rho(r_\pi)$ suggests that, as temperature increases, the analyte moves further from its initial position, increasing the probability of dissociation.

Furthermore, it is possible to note the dissociation of the 9a + TNT and 9b + TNP systems at $T = 650$ K. Full trajectory videos are included in the Supporting Information, reinforcing the argument that adsorption dissociation time is greatly reduced as temperature increases.

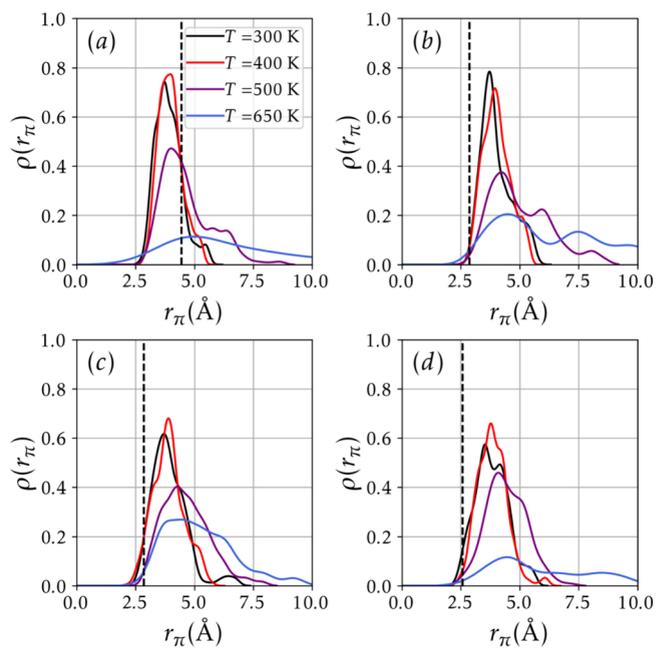

**Figure 14.** Distribution of $\rho(r_\pi)$ over the 100 ps trajectories for (a) 9a + TNT, (b) 9a + TNP, (c) 9b + TNT, and (d) 9b + TNP systems. In all panels, the vertical black dashed line represents the initial value of $r_\pi$.

Figure 15 shows the velocity and dipole autocorrelation functions estimated for the adsorbed and isolated compounds at $T = 300$ K.

These results demonstrate that both 9a and 9b compounds exhibit a noticeable shift toward higher frequencies upon adsorption of the TNT and TNP nitroaromatic compounds (NACs). While the peak positions in the VDOS and IR spectra remain largely consistent between the pristine substrates and the adsorbed complexes—reflecting the intrinsic vibrational modes of the organic framework,[86] the overall spectral shift suggests that compounds 9a and 9b are promising candidates for NAC sensing involving Raman and IR spectra.

In summary, our results suggest that changes in the electronic and vibrational properties upon NAC adsorption could be probed via electrical (I−V, impedance, conductivity), optical (fluorescence quenching), and vibrational (IR, Raman) measurements, supporting the use of these low-cost materials as promising NAC sensors.

## 4. CONCLUSIONS

In this study, the sensing capabilities of melanin-inspired compounds toward nitroaromatic compounds (NACs) were systematically investigated using density functional theory (DFT) and Born−Oppenheimer molecular dynamics (BOMD) simulations.

The results reveal that dinitro and trinitro NACs (particularly TNT and TNP) modulate the electronic, optical, and vibrational properties of the modeled systems. In general, the responses are robust across multiple adsorption relative positions.

Strong analyte−substrate interactions are noticed for these compounds, which also present a moderate estimated recovery time under mild conditions. BOMD indicates that the complexes are stable even under ambient and moderately elevated temperatures.

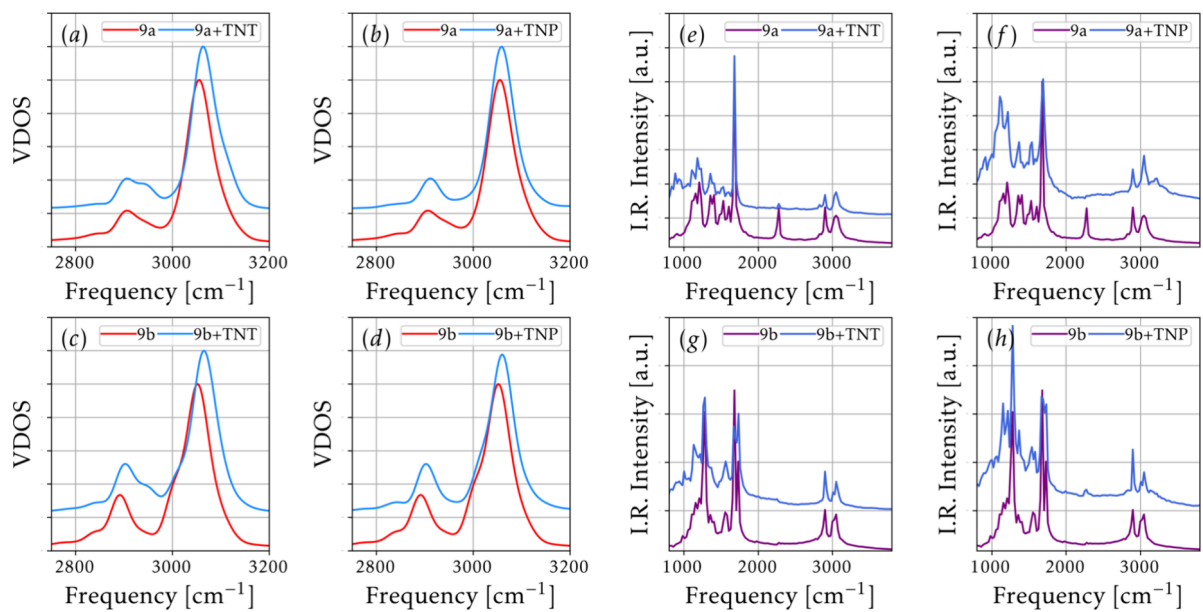

**Figure 15.** Vibrational density of states obtained from the Fourier transform of the velocity autocorrelation function for: (a) 9a + TNT, (b) 9a + TNP, (c) 9b + TNT, and (d) 9b + TNP systems. In all panels, a shift toward higher frequencies is observed upon analyte adsorption. The resulting Infrared spectrum obtained from the dipole autocorrelation function is shown in panels: (e) for 9a + TNT, (f) for 9a + TNP, (g) for 9b + TNT, and (h) for 9b + TNP.

Our results highlight the potential of melanin-inspired derivatives as suitable materials for chemiresistive and electrochemical sensors. Although the adsorbed systems exhibited notable modulation in electronic and vibrational properties, no consistent trend was observed in optical absorption shifts across all analytes. This underscores the limitation of using optical absorption alone as a sensing mechanism. Nevertheless, the bandgap reduction induced by analyte adsorption suggests a potential for luminescence-based detection strategies. In this sense, the investigation of photoluminescence and exciton dynamics represents a promising direction for the development of eumelanin-based nitroaromatic sensing platforms.

Their favorable optoelectronic and vibrational properties, combined with appropriate adsorption energies, support their use in the selective and reversible detection of nitroaromatic compounds. Consistent with experimental findings from related systems, these results position compounds 9a and 9b as promising candidates for the development of low-cost, sustainable sensor platforms, while also guiding the rational design of new bioinspired sensing materials.

## ASSOCIATED CONTENT

**Data Availability Statement**

All results in this study are reproducible using the fully optimized structures (for both isolated and adsorbed systems), available at: https://drive.google.com/drive/folders/1vFNCKKfilp1bvJVoJ5tOpIUZkpWTceHc?usp=drisve_link

**Supporting Information**

The Supporting Information is available free of charge at https://pubs.acs.org/doi/10.1021/acsomega.5c03409.

> Total density of states (DOS) of compounds 9a,b and NACs; adsorption methodologies; additional data for adsorbed systems (clusters obtained via CAFI); and results for clusters obtained via docking by aISS (PDF)
>
> BOMD NVT trajectories for 9a + TNP (MOV)
>
> BOMD NVT trajectories for 9a + TNT (MOV)

## AUTHOR INFORMATION

**Corresponding Author**

Augusto Batagin-Neto − *School of Sciences, POSMAT, São Paulo State University (UNESP), Bauru, SP 17033-360, Brazil; Department of Sciences and Technology, Institute of Sciences and Engineering, São Paulo State University (UNESP), Itapeva, SP 18409-010, Brazil;* orcid.org/0000-0003-4609-9002; Email: a.batagin@unesp.br

**Authors**

João P. Cachaneski-Lopes − *School of Sciences, POSMAT, São Paulo State University (UNESP), Bauru, SP 17033-360, Brazil; CNRS/University of Pau and the Adour Region/E2S (UPPA), Institute of Analytical Sciences and Physicochemistry for the Environment and Materials, UMR5254, 64000 Pau, France*

Felipe Hawthorne − *Department of Physics, Federal University of Paraná (UFPR), 81530-015 Curitiba, PR, Brazil; Interdisciplinary Center for Science, Technology, and Innovation (CICTI), Federal University of Paraná (UFPR), 81530-000 Curitiba, PR, Brazil;* orcid.org/0000-0002-5578-4987

Cristiano F. Woellner − *Department of Physics, Federal University of Paraná (UFPR), 81530-015 Curitiba, PR, Brazil; Interdisciplinary Center for Science, Technology, and Innovation (CICTI), Federal University of Paraná (UFPR), 81530-000 Curitiba, PR, Brazil;* orcid.org/0000-0002-0022-1319

Toby L. Nelson − *University of Tennessee, Oak Ridge Innovation Institute, Oak Ridge, Tennessee MS6173, United States*

Roger C. Hiorns − *CNRS/University of Pau and the Adour Region/E2S (UPPA), Institute of Analytical Sciences and Physicochemistry for the Environment and Materials, UMR5254, 64000 Pau, France*

Carlos F. O. Graeff − *Department of Physics and Meteorology, School of Sciences, São Paulo State University (UNESP), Bauru, SP 17033-360, Brazil*

Didier Bégué − *CNRS/University of Pau and the Adour Region/E2S (UPPA), Institute of Analytical Sciences andPhysicochemistry for the Environment and Materials, UMR5254, 64000 Pau, France;* orcid.org/0000-0002-4553-0166

Complete contact information is available at:
https://pubs.acs.org/10.1021/acsomega.5c03409

**Funding**

The Article Processing Charge for the publication of this research was funded by the Coordenacao de Aperfeicoamento de Pessoal de Nivel Superior (CAPES), Brazil (ROR identifier: 00x0ma614).

**Notes**

The authors declare no competing financial interest.

## ■ ACKNOWLEDGMENTS

The authors thank the Brazilian National Council for Scientific and Technological Development (CNPq) (grant 310390/2021-4 (AB-N)), the Coordination for the Improvement of Higher Education Personnel (CAPES) (grants 88887.817519/2023-00 CAPES-Proex and 88887.802762/2023-00 CAPES-PrInt (JPC-L)), and São Paulo Research Foundation (FAPESP) (grants 23/12686-6 (CFOG and RCH), 20/12356-8 (CFOG), 2013/08293-7, and 2019/17874-0) for financial support. Computational resources were provided by the Center for Scientific Computing (NCC/Grid-UNESP) at São Paulo State University (UNESP), the Center for Computing in Engineering and Sciences at Unicamp, the HPC Cluster Coaraci, and the Brazilian National Laboratory for Scientific Computing (LNCC) through access to the SDumont Cluster (SINAPAD/2014 project 01.14.192.00). Additional support was provided by Grand Equipement National de Calcul Intensif - Institute for Development and Resources in Intensive Scientific Computing (GENCI−IDRIS) (Grant 2022-102485).